\documentclass{article}
\usepackage{spconf,amsmath,graphicx}

\usepackage{amsmath,graphicx}
\usepackage[english]{babel}
\usepackage[latin1]{inputenc}
\usepackage{tikz}
\usepackage{url}
\newcommand{\displaycomments}

\ifdefined\displaycomments
\newcommand{\note}[1]{\textcolor{red}{\emph{#1}}}
\else
\newcommand{\note}[1]{}
\fi

\usepackage{xspace}
\usepackage{amsmath}
\usepackage{amsthm}
\usepackage{amssymb}
\usepackage{graphicx}      
\usepackage{algorithm}
\usepackage{algorithmicx}
\usepackage{cite}
\usetikzlibrary{arrows}

\DeclareMathOperator*{\maximize}{maximize}
\DeclareMathOperator*{\subj}{subject \  to}
\DeclareMathOperator*{\var}{Var}

\newtheorem{mydef}{Definition}
\newtheorem{lem}{Lemma}

\newtheorem{prop}{Proposition}

\renewcommand{\d}{\:{\rm{d}}} 
\newcommand{\E}{\mathop{\mathbb E}}

\renewcommand{\t}[1]{\mathrm{T}#1}
\newcommand{\N}{\mathcal{N}}

\usepackage{cite} 
\usepackage{stfloats}
\usepackage{amsmath}

\title{Maximum Entropy property of discrete-time  stable spline kernel}

\name{Tohid Ardeshiri and Tianshi Chen\thanks{This research has been partially supported by a research grant for junior researchers, No. 2014-5894 and the frame project grant ETT (621-2010-4301), both funded by the Swedish Research Council, and the ERC advanced grant LEARN, no. 267381, funded by the European Research Council.}}
\address{Division of Automatic Control, Department of Electrical Engineering,\\
Link\"{o}ping University, 581 83 Link\"{o}ping, Sweden,\\
email: \{tohid,tschen\}@isy.liu.se}

\begin{document}

\maketitle
\begin{abstract}
In this paper, the maximum entropy property of the discrete-time
first-order stable spline kernel is studied. The advantages of
studying this property in discrete-time domain instead of
continuous-time domain are outlined. One of such advantages is that
the differential entropy rate is well-defined for discrete-time
stochastic processes. By formulating the maximum entropy problem for
discrete-time stochastic processes we provide a simple and
self-contained proof to show what maximum entropy property
the discrete-time first-order stable spline kernel has.
\end{abstract}
\section{Introduction}
\label{sec:intro} System identification is about how to construct
mathematical models based on observed data, see e.g.,
\cite{Ljung:99}. For linear time-invariant (LTI) and causal systems,
the identification problem can be stated as follows. Consider
\begin{align}
  y(t_i) = f*u(t_i) + v(t_i), \quad i=0,1,\cdots, N
\end{align}
where $t_i,i=0,1,\cdots,N$ are the time instants at which the
measured input $u(t)$ and output $y(t)$ are collected, $v(t)$ is the
disturbance, $f(t)$ is the impulse response with $t\in\mathbb
R^+\triangleq[0, \infty)$ for continuous-time systems and
$t=t_i,\ i=0,1,\cdots$ for discrete-time systems, and $f*u(t_i)$ is
the convolution of $f(\cdot)$ and $u(\cdot)$ evaluated at $t=t_i$.
The goal is to estimate $f(t)$ as good as possible.

Recently, there have been increasing interests in system
identification community to study system identification problems
with machine learning methods, see e.g., \cite{LHO:11},
\cite{PDCDL14}. An emerging trend among others is to apply Gaussian
process regression methods for LTI, stable and causal system
identification problems, see \cite{PN10a} and its follow up papers
\cite{PCN11}, \cite{COL12a}, \cite{CACLP14}. Its idea is to model
the impulse response $f(t)$ with a suitably defined Gaussian process
which is characterized by\begin{align}
f(t)\sim\text{GP}(m(t),k(t,s)),
\end{align} where $m(t)$ is the mean function and is often set to be
zero, and $k(t,s)$ is the covariance function, also called the
kernel function in machine learning and statistics, see e.g.,
\cite{RasmussenW:06}.

The kernel $k(t,s)$ is parametrized by a hyper-parameter $\beta$
and further written as $k(t,s;\beta)$. The key issue is to design a
suitable parametrization of $k(t,s;\beta)$, or in other words, the
structure of $k(t,s;\beta)$, because it reflects our prior knowledge
about the system to be identified. Several kernel structures  have
been proposed  in the literature, e.g., the
stable spline (SS) kernel in \cite{PN10a} and the diagonal and
correlated (DC) kernel in \cite{COL12a}.

Our prior knowledge is however never complete and it is thus worth
to note Jaynes's maximum entropy rationale \cite{jaynes} to derive
complete statistical prior distributions from incomplete a priori
information.  By maximizing the entropy rate of a stochastic process
subject to constraints imposed by prior knowledge, the stochastic
process which encompasses the least assumptions about the data can
be obtained.

Interestingly, \cite{pillonetto2011} shows based on a result in
\cite{DeTL98} that for continuous-time systems, the continuous-time
first-order SS kernel (also derived by  deterministic arguments in
\cite{COL12a} and called Tuned Correlated (TC) kernel):
\begin{align}\label{eq:TC}
  k(t,s) = \min\{e^{-\beta t},e^{-\beta s}\}, \quad t,s\in\mathbb
R^+
\end{align}
has a certain maximum entropy property.

In Section \ref{sec:CTmaxEnt} the maximum entropy
property of the continuous-time kernel (\ref{eq:TC}) is briefly presented.
Then, we explain why it is worthwhile to study the maximum entropy
property for the discrete-time first order SS kernel
\begin{align}\label{eq:TC_dt}
  k(t,s) = \min\{e^{-\beta t},e^{-\beta s}\}, \quad
  t,s=t_i,i=0,1,\cdots
\end{align} and we will further elaborate our results in Section~\ref{sec:DTmaxEnt}.

\subsection{Maximum entropy property of continuous-time SS kernel}
\label{sec:CTmaxEnt}

In \cite{pillonetto2011}, the maximum differential entropy rate
\emph{continuous-time} stochastic process subject to constraints on
smoothness and bounded-input bounded-output (BIBO) stability is
sought. The definition of the differential entropy rate of a stationary \emph{continuous-time
Gaussian} process $g(t)$ with power spectrum $S(\omega)$ is adopted from \cite{DeTL98} in \cite{pillonetto2011}:
\begin{align}\label{eq:def_De}
&\overline{H}(g)=\frac{1}{4\pi}\int_{-\infty}^{+\infty}{\log\big(S(\omega)\big) \d
\omega}.
\end{align}
 We describe how smoothness and stability constraints are expressed in \cite{pillonetto2011} in separate items.

\vspace{0.2cm}
\noindent\textbf{1) Smoothness:} The smoothness constraint on the impulse responses
is addressed by using \cite[Theorem 1]{DeTL98} which suggests that the
smoothness of a signal (with some of its derivatives continuous and
bounded) can be imposed by assuming that the variances of these
derivatives are finite.
The main result in \cite[Theorem 1]{DeTL98} is given in Proposition~\ref{prop:DeNicalao} for the sake of completeness.

\begin{prop}
\label{prop:DeNicalao}\cite[Theroem 1]{DeTL98}  Let $g(t)$ be a
zero-mean bandlimited stationary Gaussian process with power
spectrum $S(\omega)=0$  for $|\omega|> B$. Given finite
$\lambda_k^2$, $k=0,1,\cdots,m$, assume that there exist real
numbers $\alpha_j$, $j=0,1,\cdots,m$ such that
$\int_{-B}^{B}\frac{\omega^{2k}}{\sum_{j=0}^m \alpha_jw^{2j}}\d\omega
= 2\pi\lambda_k^2$, $k=0,1,\cdots,m$. Under this assumption, if
there exists $S(\omega)$ that maximizes $\overline{H}(g)$ in (\ref{eq:def_De})
subject to constraints $\var[\frac{\d^k
g(t)}{\d t^k}]=\lambda_k^2$, $k=0,1,\cdots,m$, then  the spectrum is
given by $S(\omega) = \frac{1}{\sum_{j=0}^m \alpha_jw^{2j}}$. In
particular, if there is no constraints on the first $m-1$ order
derivatives, then the spectrum becomes $S(\omega) =
\frac{1}{\alpha_mw^{2m}}$.
\end{prop}
It is further claimed in  \cite{pillonetto2011} that as $B\rightarrow\infty$, the Wiener
process is the maximum differential entropy rate process among all
\emph{Gaussian} processes whose 1st-order derivatives are stationary
Gaussian processes with finite variance.

\vspace{0.2cm}
\noindent
\textbf{2) Stability:} The BIBO stability constraint on the impulse
response $f(t)$ is imposed by using a \emph{stable} time
transformation: $f(t)=g(e^{-\beta t})$ where $g(t)$ is the Wiener
process defined on $[0,1]$ and $\beta\in\mathbb R^+$. Adding this
on top of \cite[Theorem 1]{DeTL98} leads to the maximum differential
entropy rate result in \cite[Proposition 2]{pillonetto2011}
regarding the SS kernel (\ref{eq:TC}).



Deriving the maximum entropy process in continuous-time in \cite{DeTL98} and \cite{pillonetto2011} is quite involved, due to the infinite-dimensional nature of of the problem and absence of a well-defined differential entropy rate for a generic continuous-time stochastic process.

%
\subsection{Our contributions}
In this paper, we focus on discrete-time impulse responses
(stochastic processes), and provide a simple and self-contained
proof to show  the maximum entropy property of the discrete-time
first-order SS kernel (\ref{eq:TC_dt}). The advantages of working in
discrete-time domain include
\begin{enumerate}
\item The differential entropy rate is well-defined for discrete-time stochastic process.\vspace{-3mm}
\item Given a stochastic process, its finite difference process can be well-defined in discrete-time domain.\vspace{-3mm}
\item It is possible to show what maximum entropy property a zero-mean discrete-time Gaussian process with covariance function (\ref{eq:TC_dt}) has.\vspace{-2mm}
\end{enumerate}
Also, we define the discrete-time Wiener process and prove its
maximum entropy property.

\section{The Discrete-time stable spline kernel}
\label{sec:DTmaxEnt} Before deriving the maximum entropy kernels for
discrete-time  processes, we give some definitions. In the
rest of the paper the ordered index set $\mathcal{T}$ is defined as
$\mathcal{T}=\{t_i| t_0=0,t_i<t_{i+1},i=0,1,\cdots,\infty\}$ also,
the points $t_i$ in the index set $\mathcal{T}$ do not have to be
equidistant.
\begin{mydef}
The differential entropy  of a continuous random variable $X$ with density $p(x)$ is defined as
\begin{equation}
H(X)=-\int_S{p(x)\log{p(x)}\d x},
\end{equation}
where, $S$ is the support set of the random variable \cite{Cover}. \hfill $\blacksquare$
\end{mydef}
\begin{mydef}
The differential entropy rate of a real-valued discrete-time
stochastic process $\{f(t_i):\ f(t_i)\in \mathbb{R},\  t_i \in
\mathcal{T}\}$
is defined as
\begin{equation}
\label{eq:defH}
\overline{H}(f)=\lim_{n\rightarrow\infty}\frac{1}{n} H(f(t_1),f(t_2),...,f(t_n))
\end{equation}
if the limit exists \cite{Cover}.    \hfill $\blacksquare$
\end{mydef}

\begin{mydef}
The discrete-time Gaussian white noise process is a
discrete-time Gaussian process whose covariance function is
$\sigma^2\delta(t-\tau)$ \cite{Vantrees}  where $\delta (t-\tau)$ is
equal to $1$ for $t=\tau$ and $0$ otherwise. \hfill $\blacksquare$


\end{mydef}

%
In Lemma \ref{lem:WGN} we show that Gaussian white noise
process is the maximum differential entropy rate stochastic process
with constant and finite variance. The proof is an adaptation of
proof of Burg's maximum entropy theorem in \cite{Cover}.
\begin{lem}
\label{lem:WGN} The discrete-time Gaussian white noise
process is the maximum differential entropy rate stochastic process
on $\mathcal{T}$ with constant and finite variance.
\begin{proof}
First, let us formulate the  maximum differential entropy rate
problem.
\begin{equation}
 \begin{aligned}
 & \maximize_h
 & &  \overline{H}(h)\\
 & \subj
 & & \var[h(t)]=\lambda & \mbox{ for } 0<\lambda<\infty \\
 \end{aligned}
\end{equation}
where, $\text{Var}[\cdot]$ is the variance operator. In the
following, we show that $h(t)$ is a Gaussian white noise process
with variance $\lambda$.

Let $h(t_1),\ h(t_2), \cdots, \ h(t_n)$ be any stochastic process that  satisfies the constraint $\text{Var}[h(t)]=\lambda$.

Also, let $q(t_1),\ q(t_2), \cdots, \ q(t_n)$ be a Gaussian process
with the same covariance matrix as $h(t_1),\ h(t_2), \cdots, \
h(t_n)$ \footnote{Note that we are not making any assumptions
regarding the off-diagonal elements of the covariance matrix}. The
multivariate Gaussian distribution maximizes the entropy over all
$n-$dimensional  vector valued random variables under a covariance
constraint \cite{Cover}, therefore
\begin{align*}
&H\big(h(t_1),\ h(t_2), \cdots, \ h(t_n)\big)\le H\big(q(t_1),\ q(t_2), \cdots, \ q(t_n)\big).
\end{align*}
using the  the chain rule and owing to the fact that conditioning reduces the entropy we obtain
\begin{align}
&H\big(q(t_1),\ q(t_2), \cdots, \ q(t_n)\big)\nonumber\\
&=H\big(q(t_1)\big)+\sum_{i=2}^n{H\big(q(t_i)| q(t_{i-1}), q(t_{i-2}), \cdots, \ q(t_1)\big)}\nonumber\\
&\le H\big(q(t_1)\big)+\sum_{i=2}^n{H\big(q(t_i)\big)}=\sum_{i=1}^n{H\big(q(t_i)\big)}.
\end{align}
Since $q(t_1),\ q(t_2), \cdots, \ q(t_n)$ obeys a multivariate Gaussian distribution and all the diagonal entries  of the covariance matrix are equal to $\lambda$, all $q(t_i)$ are distributed according to a uni-variate Gaussian distribution with variance $\lambda$. Also, it is known that the entropy of a uni-variate Gaussian distribution depends only on its variance.  Hence, $\sum_{i=1}^n{H\big(q(t_i)\big)}=n{H\big(q(t_1)\big)}$.

Now define $q^\prime(t_1),\ q^\prime(t_2), \cdots, \ q^\prime(t_n)$ as a Gaussian white noise process where $q^\prime(t_1),\ q^\prime(t_2), \cdots, \ q^\prime(t_n)$ are identically distributed as $q(t_1)$. Since
\begin{equation}
H\big(q^\prime(t_1),\ q^\prime(t_2), \cdots, \ q^\prime(t_n)\big)=n H\big(q(t_1)\big),
\end{equation}
we obtain
\begin{align}
H\big(h(t_1),\ h(t_2), &\cdots, \ h(t_n)\big)\nonumber\\
&\le H\big(q^\prime(t_1),\ q^\prime(t_2), \cdots, \ q^\prime(t_n)\big).
\end{align}
Dividing by $n$ and taking the limit, we obtain
\begin{align}
&\lim_{n\rightarrow\infty}\frac{1}{n}H\big(h(t_1),\ h(t_2), \cdots, \ h(t_n)\big)\nonumber\\
&\le\lim_{n\rightarrow\infty}\frac{1}{n} H\big(q^\prime(t_1),\ q^\prime(t_2), \cdots, \
q^\prime(t_n)\big)=\overline{H}^*
\end{align}
where $\overline{H}^*=\frac{1}{2}\log( 2 \pi e \lambda)$ and is the differential entropy rate of the
Gaussian white noise process. Hence, the maximum differential
entropy rate stochastic process with constant and finite variance
$\lambda$ is the Gaussian white noise process with variance
$\lambda$.
\end{proof}
\end{lem}

The Wiener process $W(t)$ is a continuous-time stochastic process
which can be defined as the definite integral of continuous-time
zero-mean Gaussian white noise, has many applications in applied mathematics
and signal processing \cite{arnold}. The Wiener process can be
characterized by these properties~\cite{durrett2010}
\begin{enumerate}
\item The initial condition $W(0)=0$.\vspace{-2mm}
\item The function $W(t)$ is almost surely continuous everywhere.\vspace{-2mm}
\item $W(t)$ has independent increments with $ W(t)-W(\tau)\thicksim \N\big(0, \lambda(t - \tau)\big)$  for $0 \leq \tau < t$ , where $\N(\mu, \sigma^2)$ denotes the Gaussian distribution with mean $\mu$ and variance $\sigma^2$.\vspace{-2mm}
\end{enumerate}

In the following we will define a stochastic process, we refer to as
discrete-time Wiener process and show two of its properties in
Lemmas~\ref{lem:Wiener} and~\ref{lem:covWiener} and its maximum
differential entropy rate property in
Proposition~\ref{prop:Wiener}.
\begin{mydef}
The discrete-time Wiener process $\{f(t_i):\ f(t_i)\in \mathbb{R},\  t_i \in \mathcal{T} \}$ is characterized by these properties
\begin{enumerate}
\item $f(t_0)=0$,
\item $f(t)$ has independent increments with $ f(t_i)-f(t_j)\thicksim \N\big(0, \lambda(t_i - t_j)\big)$  for $0 \le t_j < t_i$.\hfill $\blacksquare$
\end{enumerate}
\end{mydef}
\begin{lem}
\label{lem:Wiener} The discrete-time stochastic process  $g(t)$ on
$\mathcal{T}$ is  a discrete-time Wiener process if and only if
$g(t_0)=0$ and
\begin{align}\label{eq:wiener}
g(t_n)=\sum_{i=1}^{n}{h(t_i)}\sqrt{t_i-t_{i-1}},&& \  n\geq1
\end{align}
where $h(t)$ is the zero-mean Gaussian white noise process.
\begin{proof}
First, we prove the necessary part. That is, we show if $g(t)$ is a
discrete-time Wiener process, it can be expressed in the form of
(\ref{eq:wiener}). Let $w(t_i)\triangleq
\frac{g(t_i)-g(t_{i-1})}{\sqrt{t_i - t_{i-1}}}$ for
$i\in\mathbb{N}$. Since $ g(t_i)-g(t_{i-1})\thicksim \N\big(0,
\lambda(t_i - t_{i-1})\big)$ we have
\begin{equation}
w(t_i)\thicksim \N\big(0, \lambda\big).
\end{equation}
Also, since $g(t)$ has independent increments it follows that $w(t)$ is a
discrete-time zero-mean Gaussian white noise process. Also, from the
definition of $w(t)$ we have 
\small{\begin{equation}\nonumber
g(t_n)={w(t_n)}\sqrt{t_n-t_{n-1}}+g(t_{n-1})=\sum_{i=1}^{n}{w(t_i)}\sqrt{t_i-t_{i-1}}.
\end{equation}}\normalsize
Now we prove the sufficient part, i.e., the stochastic process
(\ref{eq:wiener}) is a discrete-time Wiener process. Since Gaussian
processes are closed under linear operations \cite{RasmussenW:06},
$g(t)$ is a Gaussian process. Also $\E[g(t_n)]=\E\left[\sum_{i=1}^{n}{h(t_i)}\sqrt{t_i-t_{i-1}}\right]=0$.
Furthermore, let $0 \le t_j < t_i$,
\begin{align}
&\text{Var}[g(t_i)-g(t_j)]=\E\left[\left(\sum_{r={j+1}}^{i}{h(t_r)\sqrt{t_r-t_{r-1}}}\right)^2\right]\nonumber\\
&=\sum_{r={j+1}}^{i}{\lambda ({t_r-t_{r-1}})}=\lambda (t_i-t_j).
\end{align}
Therefore, $ g(t_i)-g(t_j)\thicksim \N\big(0, \lambda(t_i -
t_j)\big)$  for $0 \le t_j < t_i$. Since $h(t)$ is a Gaussian white
noise process, the increments of $g(t)$ are independent and the proof
follows.
\end{proof}

\end{lem}
\begin{lem}
\label{lem:covWiener}
The covariance of the discrete-time Wiener process is given by
\begin{equation}
\mathbb{V}[g(t_i),g(t_j)]=\lambda \min\{t_i,t_j\}\ \ \   \text{ for } \ t_i,t_j \in \mathcal{T}
\end{equation}
where, $\lambda$ is the variance of the underlying Gaussian white noise process.
\begin{proof} Since $\E[g(t_i)]=\E[g(t_j)]=0$, then 
\begin{align}
&\mathbb{V}\left[g(t_i),g(t_j)\right]=\E\left[g(t_i)\cdot g(t_j)\right]\nonumber\\
&=\E\left[\left(\sum_{r=1}^{i}{h(t_r)\sqrt{t_r-t_{r-1}}}\right)\left(\sum_{s=1}^{j}{h(t_s)\sqrt{t_s-t_{s-1}}}\right)\right]\nonumber\\
&=\sum_{q=1}^{\min(i,j)}{\lambda({t_q-t_{q-1}})}=\lambda \min\{t_i,t_j\}
\end{align}
\end{proof}
\end{lem}
In Proposition \ref{prop:Wiener} the maximum differential entropy
rate property of the discrete-time Wiener process when the index set is unbounded from above is studied. For
the continuous-time case, the smoothness constraint on the
stochastic process is expressed by constraining  the variance of its
first-order derivative  to be constant and finite, see Proposition
\ref{prop:DeNicalao}. Due to the absence of derivative for the
Wiener process we use the variance of the finite difference of the
discrete-time stochastic process defined below.
\begin{mydef} The finite difference of a discrete-time stochastic process $f(t)$ on $\mathcal{T} $, at $t_i\in\mathcal{T}$ is an expression of the form
\begin{equation}
\Delta [f] (t_i) \triangleq f(t_{i+1})-f(t_{i}).
\end{equation}
\hfill $\blacksquare$
\end{mydef}
\begin{prop}
\label{prop:Wiener} The discrete-time Wiener process is the maximum
differential entropy rate stochastic process on $ \mathcal{T}$ with
$t_\infty=\infty$ and $\lim_{n \rightarrow \infty}\frac{1}{n}\sum_{i=1}^n\log\sqrt{t_{i}-t_{i-1}}<\infty$  such that its value at origin is zero and is
zero-mean and variance of its finite difference at all
$t_i\in\mathcal{T}$ is proportional to the time increment
$t_{i+1}-t_{i}$ and $ t_{i+1}-t_{i}$ is bounded from below by a
positive number. That is, the discrete-time Wiener process is the
optimal solution to the problem:
\begin{equation}
\label{eq:maxEntWiener}
 \begin{aligned}
 & \maximize_g 
 \quad \overline{H}(g)\\
 & \subj 
  \quad g(t_0)=0\\
&\E[g(t)]=0\\
 &  \var \left[\Delta[g](t_i)\right]=\lambda (t_{i+1}-t_{i}), \  \ \ \lambda>0, \\
& t_{i+1}-t_{i}\geq \delta>0,\ \ \ \  i=0,1,\cdots,\infty
\end{aligned}
\end{equation}

\begin{proof}
Let  $g(t)$ be any discrete-time stochastic process on
$\mathcal{T}$. Now we define the stochastic process $w(t)$ as
\begin{equation}
\label{eq:defw}
w(t_{i+1})\triangleq\frac{\Delta[g](t_i)}{\sqrt{t_{i+1}-t_{i}}}.
\end{equation}
Therefore, $\E[w(t)]=0$ and the variance of the finite difference of $g(t)$ obeys, $\var\left[{\Delta[g](t_i)}\right]=\E\left[w(t_{i+1})^2\right]({t_{i+1}-t_{i}}).$
So the third constraint in the maximization problem
\eqref{eq:maxEntWiener} can be written as
$\text{Var}[w(t_{i+1})]=\lambda$. Also, from \eqref{eq:defw} we
have\begin{align}
g(t_{n+1})&=g(t_{n})+w(t_{n+1})\sqrt{t_{n+1}-t_{n} } \\
& =\sum_{i=1}^{n+1}{w(t_i)\sqrt{t_i-t_{i-1}}}.
\end{align}
Let $\mathcal{G}\triangleq[g(t_1), \cdots ,g(t_n)]^\t{}$ and
$\mathcal{W}\triangleq[w(t_1), \cdots ,w(t_n)]^\t{}$. We have
$\mathcal{G}=A\mathcal{W}$ where $A$ is a lower triangular
non-singular matrix independent of $\mathcal{W}$ and $\mathcal{G}$.
Using~\eqref{eq:defH} and \cite[Corollary to Theorem 8.6.4]{Cover}
we obtain
\begin{equation}
H(\mathcal{G})=H(\mathcal{W})+\log |A|
\end{equation}
Therefore, it is sufficient to maximize the differential entropy
rate of the underlying stochastic process $w(t)$ such that the
variance of $w(t)$ is constant and $\E[w(t)]=0$. Using Lemma~\ref{lem:WGN} $w(t)$
turns out to be a zero-mean Gaussian white noise process. Consequently, using
Lemma~\ref{lem:Wiener} the maximum differential entropy rate
stochastic process $g(t)$ turns out to be the discrete-time Wiener
process.
\end{proof}
\end{prop}

\noindent\textbf{Remark 1.} The assumption $\lim_{n \rightarrow \infty}\frac{1}{n}\sum_{i=1}^n \log \sqrt{t_{i}-t_{i-1}}\allowbreak < \infty$ is not restrictive. 
For example, the assumption is trivially satisfied for uniform sampling where  $t_i-t_{i-1}=T_s>0$, $i=1,...,\infty$. 
When the time increment $ t_{i+1}-t_{i}$
is not bounded from below, $\lim_{n\rightarrow \infty}\frac{1}{n}
\log|A|$ becomes infinite and thus the differential entropy rate is
not defined. In this case, the discrete-time Wiener process is the
maximum differential entropy stochastic process on the finite
segment $\{t_0,\cdots,t_n\}\subset \mathcal{T}$ in the following
sense: for any $n\in\mathbb N$, it optimizes the maximum
differential entropy problem
\begin{equation}
\label{eq:maxEntWiener_reduced}
 \begin{aligned}
& \maximize_g
 \quad H(g(t_1),\cdots, g(t_n))\\
& \subj
  \quad g(t_0)=0\\
&\E[g(t)]=0\\
& \var\left[\Delta[g](t_i)\right]=\lambda (t_{i+1}-t_{i}),\ \  \lambda>0   \\
& \quad i=0,1,\cdots,n-1
\end{aligned}
\end{equation}


Before proceeding to the maximum differential entropy  property of
the discrete-time stable spline kernel (\ref{eq:TC_dt}) in
Proposition \ref{prop:DTstablespline}, we introduce the concept of
reverse ordered index set of $\mathcal T=\{t_0,t_1,\cdots,t_n\}$
which is defined as the ordered index set $\{t_n,\cdots, t_1,t_0\}$
and denoted by $\overline{\mathcal T}$.

\begin{prop}
\label{prop:DTstablespline} Let $g(\tau)$ denote a zero-mean
discrete-time stochastic process defined on an ordered index set
$\{\tau_i|\tau_0=0,\tau_\infty=1, 0< \tau_i< \tau_j< 1,\ 0< i <j <
\infty\}$. Now consider a finite segment of $g$ with index set
$\mathcal T_g =\{\tau_i|\tau_0=0,\tau_n<1, 0< \tau_i< \tau_j<t_n,\
0< i <j < n\}$. 
Then for any $n\in\mathbb N$, the zero-mean Gaussian process with
covariance function (\ref{eq:TC_dt}) is the solution to the maximum
differential entropy problem:
\begin{equation}
\label{eq:maxEntSS_dt}
\begin{aligned}
& \maximize_f \quad  H(f(t_0),\cdots,f(t_{n-1}))\\
& \subj   \quad f(t)=g(e^{-\beta t}),\ \beta>0,
t\in\overline{\mathcal T_g} ,\\
&  g(\tau_0)=0,\\
&\E\left[g(\tau)\right]=0,\\
& \var\left[\Delta[g](\tau_i)\right]=\lambda
(\tau_{i+1}-\tau_{i}), i=0,1, \cdots,n-1
\end{aligned}
\end{equation}
\begin{proof} Note that \begin{align}
H(&f(t_0),\cdots,f(t_{n-1})) = H(g(e^{-\beta
t_0}),\cdots,g(e^{-\beta t_{n-1}}))\nonumber\\&=
H(g(\tau_{n}),\cdots,g(\tau_1)).
\end{align}
which implies that (\ref{eq:maxEntSS_dt}) is equivalent to the
maximum entropy problem (\ref{eq:maxEntWiener_reduced}). From
Proposition \ref{prop:Wiener} and Remark 1, the optimal solution to
(\ref{eq:maxEntWiener}) is the discrete-time Wiener process $g(t)$
which is zero-mean Gaussian and has covariance function
$\lambda\min\{t,s\}$, $t,s\in\{\tau_i|\tau_0=0,\tau_\infty=1, 0<
\tau_i< \tau_j< 1,\ 0< i <j < \infty\}$ (from Lemma
\ref{lem:covWiener}). As a result, the optimal solution to
(\ref{eq:maxEntSS_dt}) is the zero-mean Gaussian process induced by
$f(t)=g(e^{-\beta t})$ which has the covariance function
$\lambda\min\{e^{-\beta t},e^{-\beta s}\}$.
\end{proof}
\end{prop}

\section{conclusion}
The maximum entropy property of the first-order discrete-time stable spline
kernel for identification of LTI stable and causal systems are
studied. By formulating the maximum entropy problem for
discrete-time stochastic processes we provide a simple and
self-contained proof to show  the maximum entropy property of the
discrete-time first-order stable spline kernel. Also, we define the
discrete-time Wiener process and prove its maximum entropy property.

\bibliographystyle{IEEEbib}
\bibliography{taref2014}
\end{document}